\documentclass[12pt]{article}
\usepackage{epsfig}   
\newcommand{\be}{\begin{equation}}   
\newcommand{\ee}{\end{equation}}   
\newcommand{\beqn}{\begin{eqnarray}}   
\newcommand{\eeqn}{\end{eqnarray}}
\newcommand{\mbf}[1]{\mbox{\boldmath $#1$}} 
\begin{document}
DESY 00-077                           \hfill ISSN 0418-9833 \\
\begin{center}   
\begin{Large}   
\vspace*{0.5cm}  
{\bf Properties of a family of $n$ reggeized gluon states in multicolour QCD}\\ 
\end{Large}   
\vspace{0.5cm}   
G.P. Vacca \footnote{
Supported by the Alexander von Humboldt Stiftung.\\
E-mail: vacca@bo.infn.it}\\
II. Inst. f. Theoretische Physik,    
Univ. Hamburg, Luruper Chaussee 149, D-22761 Hamburg\\  
\end{center}   
\vspace*{0.75cm} 
\begin{abstract}
\noindent
A general relation between families of $(n+1)$ gluon and $n$ gluon eigenstates
of the BKP evolution kernels in the multicolour limit of QCD is derived.
It allows to construct an $(n+1)$ gluon eigenstate if an $n$ gluon eigenstate
is known; this solution is Bose symmetric and thus physical for even $n$.   
A recently found family of odderon solutions corresponds to the
particular case $n=2$.
\end{abstract}
\section{Introduction}
The need to unitarize the scattering amplitudes in QCD obtained in LLA
in the Regge (small $x$) limit has been felt since the first derivation
of the BFKL Pomeron \cite{BFKL}.
To achieve this very difficult task different approaches have been addressed.
One of the method, used to go beyond the two-gluon ladder approximation,
consists in investigating the solutions of the BKP equations for
multi-reggeized-gluon compound states \cite{BKP}.
Due to their rather high complexity they have been mostly analyzed 
in the large-$N_c$ limit and it has been found that, if they are casted into
the Hamiltonian form, remarkable symmetry properties \cite{Lintegr,LFKchain}
appear.
The existence of integrals of motion 
\cite{Lintegr} and the duality symmetry \cite{Ldual} provide powerful tools
in analysing the spectrum which characterizes the behaviour of the amplitudes
as a function of the center of mass energy.
Another line of research
investigates the transition between states with different numbers of gluons
\cite{B,BW,BL,BE}. In this letter, however, we shall restrict ourself
to consider fixed number of gluon states.

Recently the three gluon system (odderon \cite{NICOL}) has been intensively
studied and 
after several variational studies, an eigenfunction of the integral of motion
\cite{Lintegr} with the odderon intercept slightly below one was constructed
by Janik and Wosiek \cite{JW} (see also \cite{Ldual}) and subsequently
verified by Braun et al \cite{BGN}.
From the phenomenological side, a possible signature of the odderon in
deep inelastic scattering at HERA has been investigated by several authors,
and also the coupling of the odderon to the 
$\gamma^* \to \eta_c$ vertex has been given \cite{Kwie}.

Successively, a new branch of odderon states with an intercept up exactly
to one
was found \cite{BLV}. The first analysis was based on the discrete
symmetry structure of the pomeron $\to$
two-odderon vertex, obtained from an analysis of the
six-gluon state \cite{BE}, and the bootstrap property of the BFKL kernel
which encodes the gluon reggeization property.
At the same time, this solution was also obtained using the duality symmetry
of the three gluon Hamiltonian. 
It was also noted that this new branch of states revealed a non
zero coupling to the $\gamma^* \to \eta_c$ vertex
contrary to the those found previously. 

In this letter we study in general a system of $n$ interacting reggeized
gluons, in topthe large-$N_c$ limit, and show how some eigenstates with definite
symmetry properties in the configuration space can be constructed if some
$(n-1)$ gluon eigenstates with opposite symmetry property are known.
The main ingredients are: the discrete symmetry of the $n$
gluon Hamiltonian (evolution kernel) and the bootstrap relation
of gluon reggeization.
The construction procedure is valid for any $n>2$, but it turns out that it
is a nilpotent operation which means that it cannot be iterated twice.
The symmetry structure of the solutions, which can be constructed, is
discussed in details.
The recently found odderon solution \cite{BLV} is  related to the
specific $n=3$ case.
Due to the requirement that the full eigenstates should be Bose symmetric,
one sees that only the solutions obtained for an odd number of gluons are
physical, while for the even case they are merely solutions of the integrable
system defined on the transverse space only, without considering the symmetry
properties in the colour space.

\section{The BFKL and BKP equations in LLA}
In this section we give a brief review of the equations describing the
dynamics of the reggeized gluons in the LLA in QCD.
Let us start from the Schr\"odinger-like BFKL equation \cite{BFKL}
describing the two reggeized gluons compound state,
\be
K_2^{(R)} \otimes \psi_E = E \, \psi_E, \quad
K_2^{(R)}=K_{ij}^{(R)}= -\frac{N_c}{2}\left( \tilde{\omega}_i+
\tilde{\omega}_j \right) - \lambda_R V_{ij}.
\label{bfkleq}
\ee
Here $R$ labels the colour representation of the two gluon state and in the
singlet and octet channel one has respectively $\lambda_1=N_c$ and
$\lambda_8=N_c/2$. The symbol $\otimes$ denotes an integration in the
transverse space,
while $E=-\omega=1-j$ where $j$ is the complex variable which describes the
singularities of the $t$ channel partial waves, dual in the high energy limit
to the center of mass energy $s$ in the Mellin transform sense.

In the LLA the gluon trajectory (scaled by $N_c/2$) is given by the well
known expression
\be
\tilde{\omega}_i=\tilde{\omega}(k_i)=- \int d^2l \,
\frac{k_i^2}{l^2 (k_i-l)^2}, \quad c=\frac{g^2}{(2\pi)^3}
\ee
and the interaction term defined by its action
\be
V_{ij} \otimes \psi \, (k_i,q-k_i) = c \int d^2l \,
\left[ \frac{l^2}{k_i^2 (k_i-l)^2} + \frac{(q-l)^2}{(q-k_i)^2 (k_i-l)^2}
- \frac{q^2}{k_i^2 (q-k_i)^2} \right] \psi(l,q-l)
\ee
and $q=k_i+k_j$.

In the construction of the BFKL kernels a very important assumption
is the gluon reggeization property which can be verified with the self
consistent bootstrap relation, which guarantees that production amplitudes
with the gluon quantum numbers in their $t$ channels, used for the
construction of the absorptive part, are characterized by just a single
reggeized gluon exchange (in leading and next-to-leading orders).
The bootstrap relation can be written in terms of the gluon trajectory and
interaction terms.
It is convenient to use a slightly different form of the interaction term,
which acts on the so called amputated function space (with the propagators
removed). We shall denote this operator by $\bar{V}_{ij}$, with its form
explicitely given by
\beqn
&& \bar{V}_{ij} \otimes \phi \, (k_i,q-k_i) =
\int d^2 k_i' \, \bar{V}(k_i, k_j| k_i',k_j') \phi(k_i',k_j') \\ \nonumber
&=&  c \int d^2 k_i' \,
\left[ \frac{k_i^2}{k_i'^2 (k_i-k_i')^2} + \frac{(q-k_i)^2}{(q-k_i')^2 
(k_i-k_i')^2}
- \frac{q^2}{k_i'^2 (q-k_i')^2} \right] \phi(k_i',q-k_i')
\eeqn

Therefore for the LLA case the bootstrap condition can be written as
\be
\omega(q)-\omega(k_i)-\omega(k_j)=\bar{V}_{ij} \otimes 1,
\ee
where the constant $1$ is the wave function which can be conveniently obtained
after rescaling any function depending only on $q = k_i+k_j$.
Let us note that it is crucial in the bootstrap relation that the two gluons
are located at the same point in the transverse coordinate plane space.
In the following we will use a more compact notation, involving directly the
full kernels for the amputated functions (distinguished with a bar
from the non amputated case),
\be
\bar{K}_{ij}^{(8)}\otimes 1=-\omega(q), \quad 
\bar{K}_{ij}^{(1)}\otimes 1=-2 \omega(q)
+\omega(k_i)+\omega(k_j),
\label{bootstrap2}
\ee
valid for the octet and singlet channel, respectively.

The 2-gluon kernel (\ref{bfkleq}) in the singlet channel has been
investigated in general in the coordinate representation.
Using complex coordinates, two important
properties, the holomorphic separability and the invariance under the M\"obius
transformation were shown. In particular, the
solutions of the homogeneous BFKL pomeron equation belong to irreducible
unitary representations of the M\"obius group, and are eigenstates of its
Casimir operator:
\be
E^{m,\widetilde{m}}(\rho_{i0},\, \rho_{j0})=\left(
\frac{%
\rho _{ij}}{\rho _{i0}\rho _{j0}}\right) ^{m}\left( \frac{\rho
_{ij}^{*}}{%
\rho _{i0}^{*}\rho _{j0}^{*}}\right) ^{\widetilde{m}}\,,
\ee
where 
$m=\frac{1}{2}+i\nu +\frac{n}{2}\,,\,\,\widetilde{m}=\frac{1}{2}+i\nu
-\frac{n%
}{2}$ are conformal weights belonging to the basic series of the unitary
representations of the  M\"obius group, $n$ is the conformal spin and
$d=1-2i\nu $ is the anomalous dimension of the operator
$O_{m,\widetilde{m}}(\mbf{\rho }_{0})$ describing
the compound state \cite{Lcft}.
The corresponding eigenvalues of the BFKL kernel are given by
\be
\chi(\nu,n)=\frac{N_c\alpha_s}{\pi}
\left( \psi(\frac{1+|n|}{2}+i\nu)
+\psi(\frac{1+|n|}{2}-i\nu) -2\psi(1)\right) \, .
\ee
When more than 2 reggeized gluons in the $t$ channel are considered,
the corresponding scattering amplitudes obtained in LLA are described by
the BKP equation \cite{BKP} which can be viewed as a quantum mechanical
n-body problem with a Hamiltonian describing the dynamics of the 
pairwise interaction of the reggeized gluons.
The need to study such $n$-gluon states is connected with the problem
of finding a way to restore unitarity in the BFKL resummation approach.

The odderon \cite{NICOL}, in this context, corresponds to a special
case of the BKP equations, and represents a three-body problem.
The kernel contains terms corresponding to the gluon trajectories
and the ``interaction'' terms due to real gluon emission:
\be
K_n = - \frac{N_c}{2} \sum_i \, \tilde{\omega_i} + \sum_{i<j} T_i T_j \, V_{ij}
= \frac{1}{N_c} \sum_{i<j} T_i T_j \, K_{ij}^{(1)},
\label{bkpeq}
\ee
here the sums run up to the number of reggeized gluon $n$ and $T_j$ is
the colour operator of the corresponding gluon.
The second equality in (\ref{bkpeq}) is valid for a system of gluons
in a global colour singlet state and has the advantage to be written
in terms of the BFKL pomeron kernel.

In the multicolour limit ($N_c \to \infty$) the dominant colour structure
is planar, leading to a cylindric topology of the interactions. Each
two neighbouring gluons will be in a colour octet state.
Therefore one can obtain a simpler kernel
\be
K_n^{\infty} = \frac{1}{2} \left[ K_{12}^{(1)} + K_{23}^{(1)}
+ \cdots + K_{n1}^{(1)} \right].
\label{bkpnlarge}
\ee
The kernel $K_n^{\infty}$ has some trivial and also non trivial symmetries
\cite{Lintegr,LFKchain,Ldual}.

First, it is clearly invariant under a shift along the cylinder (rotation),
generated by the operator $R_n$, i.e. $[K_n^{\infty},R_n]=0$ with $(R_n)^n=1$.

In the coordinate representation, using complex coordinates the kernel
$K_n^{\infty}$ can be written as a complicated pseudo-differential operator
which:

(a) is holomorphic separable,
which means $K_n^{\infty}=\frac{1}{2}(h_n + h_n^*)$;

(b) has $(n-1)$ non trivial integrals of motion, represented by
the following operators $q_r$, such that $[q_r,h_n]=0$, $[q_r,q_s]=0$ (plus similar
relations for the antiholomorphic sector),
\be
q_r=\sum_{i_1<i_2< \cdots < i_r} \rho_{i_1 i_2}\rho_{i_2 i_3} \cdots
\rho_{i_r i_1}\, p_{i_1}p_{i_2} \cdots p_{i_r},
\ee
where $\rho_{i j}=\rho_i -\rho_j$ and $p_j=i \partial_j$. In particular,
$q_2=M^2$ is the Casimir of the M\"obius group.

(c) $K_n^{\infty}$ is invariant under a transformation $D_n$, called duality,
defined by $\rho_{i-1,i} \to p_i \to \rho_{i,i+1}$ on the cylinder combined
with the reversed order of operator multiplication. It can be viewed as a kind
of supersymmetry since $(D_n)^2=R_n$. An integral equation for the duality has
been given in the general case and also a differential form for the three
gluon case has been given.

A lot of effort has been spent in the last years to analyze the $n=3$ case,
in particular the odderon state, with a full symmetric wave function in the
configuration space. 

Lipatov suggested to take advantage of the integral of motion $q_3$
to search for conformal invariant eigenstates of the holomorphic part
of the kernel $h_3$ (and the same for the antiholomorphic sector).
The condition that the total full symmetric eigenstates of $K_3$, 
written in factorized form, is single-valued imposes a non trivial constraint
on the spectrum of this family of solutions.
This work was carried out by Janik and Wosiek \cite{JW}, who found an
expansion for a family of solutions with a discrete imaginary $q_3$
operator eigenvalue
and a maximum eigenvalue $E_0=0.16478 (9 \alpha_s)/(2 \pi)$ which
corresponds to the intercept slightly below $1$. This solution has been
verified with the help of variational calculations \cite{BGN}.
Also WKB analyses of odderon states have been performed \cite{Korch,Ldual},
and an agreement with the above picture was found.

Recently another set of odderon solutions has been obtained \cite{BLV},
characterized by a spectrum with a maximum intercept at $1$ and zero
$|q_3|^2$ eigenvalue.  
The peculiar symmetry structure of these eigenstates was suggested by
pomeron$ \to$ two odderon vertex,  which came out from the
study of the six gluon amplitude, and
by the impact factor $\Phi_{\gamma \to \eta_c}$,
to which this odderon states couple contrary to the previously found
eigenstates.
This set of eigenstates is written in a very simple form in the amputated
version (propagators removed), as a cyclic sum of three amputated
BFKL pomeron odd eigenstates (odd conformal spin) where two gluons have
the same transverse coordinate.
Two possible derivations of such states have been given. One is based on the
property of gluon reggeization by means of the bootstrap condition,
which in this context (the three gluons have $d_{abc}$ colour structure)
can be seen as the reggeization of the $d$-reggeon, being
the even signature partner of the gluon which belongs to the symmetric
octet representation for a two-gluon compound state.
The other derivation is based on the duality transformations of an already
known solution with different symmetry properties. 

It is still not clear if other 3-gluon states with the odderon quantum number
exist. There are some indications that the recently found set of
solutions may lie at the tail of a family of eigenstates with a continuous
$|q_3|^2$ eigenvalue. This problem, however, will not be addressed here.

The BKP equations for $n$ gluons have been mostly studied in the
large-$N_c$ limit, which restricts the domain of applicability
in the study of the true QCD dynamics ($N_c=3$).
It is clear in fact that non planar colour structures can be important; for
example in the four gluon case, even if hard to investigate,
there is a feeling that the full interaction may
lead to some states with intercept a little above the value corresponding
to the two non interacting BFKL pomeron configuration.

Nevertheless taking the limit $N_c \to \infty$ is useful for understanding
some features of small-$x$ QCD dynamics.
In such a limit the BKP equations possess a rich structure,
already partially emerged at the $3$ gluon level. In fact
they describe completely integrable systems,
equivalent to XXX Heisenberg spin model, although up till now very little is
known about the properties of their solutions.  
In the following some very particular sets of eigenstates
of the $n$-gluon system will be discussed. 
\section{A family of $n$ reggeized gluon states in multicolour QCD}

We shall start from considering the $n$-gluon eigenstates of the $|q_n|^2$
operator with formally zero eigenvalue\footnote
{Since one is dealing with distributions, this depends on the space of test
function chosen.}, 
$ |\rho_{12}\rho_{23} \cdots
\rho_{n1}|^2 \, \partial_1^2 \partial_2^2 \cdots \partial_n^2 \, E_n=0$. 
If one considers the amputated function $\varphi_n=
\partial_1^2 \partial_2^2 \cdots \partial_n^2 \, E_n$, it is therefore possible
to satisfy the above constrain for a general form
$\varphi_n=\sum_i \delta^2(\rho_{i,i+1}) g_i$,
where the sum runs over the cyclic permutations.
In particular, choosing a particular form of $g_i$, we write
\be
\varphi_n(k_1,k_2,\cdots,k_n)=
\sum_{i=0}^{n-1}  (R_n)^i c_i \, \varphi_{n-1}(k_1+k_2,k_3,\cdots,k_n),
\label{ampuans}
\ee
which, for $E_n$ in momentum representation, corresponds to
 
\be
E_n(k_1,k_2,\cdots,k_n)=
\sum_{i=0}^{n-1}  (R_n)^i c_i \frac{(k_1+k_2)^2}{k_1^2 k_2^2}
E_{n-1}(k_1+k_2,k_3,\cdots,k_n).
\label{nonampu}
\ee
Here the rotation operator $R_n$ has been used to perform a sum over the cyclic
permutations, and $c_i$ are weights which have to be determined in order
to obtain an eigenfunction of the kernel $K_n^\infty$.
It can be seen as a generalization of the form of the odderon states \cite{BLV}
for the case $n=3$. We shall see that one is also forced to
require $E_{n-1}$ to be an eigenstate of $K_{n-1}^\infty$ with different 
symmetries properties, depending on $n$.

Let us study the action of the $n$ gluon BKP kernel on the ansatz
given in Eq. (\ref{ampuans}), since it is convenient to work with the
amputated form. In particular let us isolate just the first term in the
cyclic sum and act on it with $\bar{K}_n^\infty(1,2,\cdots,n)$,
where the gluons on which the kernel acts are explicitely indicated.
It is useful to write
\be
\bar{K}_n^\infty(1,2,\cdots,n)  =
\frac{1}{2} \left( \bar{K}_{12}^{(1)} + \bar{K}_{1n}^{(1)}
- \bar{K}_{2n}^{(1)} \right) + \bar{K}_{n-1}^\infty(2,3,\cdots,n),
\label{decomp} 
\ee
where from the $n$-gluon kernel a $(n-1)$-gluon subkernel is extracted,
as can be easily checked looking at (\ref{bkpnlarge}).

Let us also make use of an integral operator, introduced by Bartels
\cite{Bart80} to describe the elementary transition between $2$ and $3$
reggeized gluons in LLA.
It can be defined with the help of the integral kernel
\be
W(k_1,k_2,k_3|k_1',k_3')=\bar{V}(k_2,k_3|k_1'-k_1,k_3')
- \bar{V}(k_1+k_2,k_3|k_1',k_3'),
\ee
which is symmetric under the exchange of the left and right gluon momenta
$(1,3)$.
Therefore one can write the following relations for the action on the function
$g = \varphi_{n-1}(k_1+k_2,k_3,\cdots,k_n)$:

(a) the last term in (\ref{decomp}) gives

\beqn
&& \bar{K}_{n-1}^\infty(2,3,\cdots,n) \otimes g = \nonumber \\
&& \int \{d^2 k_i'\} \bar{K}_{n-1}^\infty
(k_2,k_3,\cdots,k_n|k_2',k_3',\cdots,k_n') \,
\varphi_{n-1}(k_1'+k_2',k_3',\cdots,k_n')= \nonumber \\
&&\int \{d^2 k_i'\} \bar{K}_{n-1}^\infty
(k_1+k_2,k_3,\cdots,k_n|k_2',k_3',\cdots,k_n')
\, \varphi_{n-1}(k_2',k_3',\cdots,k_n') + \nonumber \\
&& 
 \Bigl[ \omega(k_1+k_2) -\omega(k_2) \Bigr] \,
\varphi_{n-1}(k_1+k_2,k_3,\cdots,k_n) + \nonumber \\
&& \frac{1}{2} \int \{d^2 k_i'\}\, \Bigl[ W(k_1,k_2,k_3|k_1',k_3') 
+  W(k_1,k_2,k_n|k_1',k_n') \Bigr] \,
\varphi_{n-1}(k_1',k_3',\cdots,k_n'); 
\eeqn

(b) the first term inside the parenthesis in (\ref{decomp})
after applying the bootstrap relation in (\ref{bootstrap2}), gives

\be
\frac{1}{2}\bar{K}_{12}^{(1)} \otimes g =
\frac{1}{2} \Bigl[\omega(k_1)+\omega(k_2) - 2\omega(k_1+k_2)\Bigr]
\varphi_{n-1}(k_1+k_2,k_3,\cdots,k_n);
\ee

(c) and finally the remaining terms in (\ref{decomp}) act in the following way
\beqn
&&\frac{1}{2} \Bigl[ \bar{K}_{1n}^{(1)} -\bar{K}_{2n}^{(1)} \Bigr] \otimes
g = \nonumber \\
&&\frac{1}{2} \Bigl[ \omega(k_2) - \omega(k_1) \Bigr]
\varphi_{n-1}(k_1+k_2,k_3,\cdots,k_n) + \nonumber \\
&& \frac{1}{2} \int \{d^2 k_i'\}\, W(k_2,k_1,k_n|k_2',k_n')
 \varphi_{n-1}(k_2',k_3',\cdots,k_n') - \nonumber \\
&& \frac{1}{2} \int \{d^2 k_i'\}\, W(k_1,k_2,k_n|k_1',k_n') \,
\varphi_{n-1}(k_1',k_3',\cdots,k_n').
\eeqn

Let us note that in each integral $\{d^2 k_i'\}$ indicates the
necessary measure, which form should be clear from the context.
Collecting all the pieces, after some cancellations, one can finally write
\beqn
&&\bar{K}_n^\infty
(1,2,\cdots,n) \otimes g =
\nonumber \\
&&\int \{d^2 k_i'\} \bar{K}_{n-1}^\infty
(k_1+k_2,k_3,\cdots,k_n|k_2',k_3',\cdots,k_n')
\, \varphi_{n-1}(k_2',k_3',\cdots,k_n') + \nonumber \\
&& \frac{1}{2} \int \{d^2 k_i'\}\,  W(k_1,k_2,k_3|k_1',k_3') 
\varphi_{n-1}(k_1',k_3',\cdots,k_n') + \nonumber \\
&&\frac{1}{2} \int \{d^2 k_i'\}\,  W(k_2,k_1,k_n|k_2',k_n') \,
\varphi_{n-1}(k_2',k_3',\cdots,k_n').
\label{collect}
\eeqn
At this point we shall require for $\varphi_{n-1}$ the following properties:
(1) $\bar{K}_{n-1}^\infty \, \varphi_{n-1}= \chi \, \varphi_{n-1}$ and
(2) $R_{n-1} \, \varphi_{n-1}= r_{n-1} \, \varphi_{n-1}$, i.e. it is chosen
to represent a $(n-1)$-gluon eigenstate with definite symmetry properties.

Taking into account these properties in (\ref{collect}) together with the
previously mentioned symmetry of the $W$ operator, one can write the following

\beqn
&&\bar{K}_n^\infty
(1,2,\cdots,n) \otimes g =
\chi \, \varphi_{n-1}(k_1+k_2,k_3,\cdots,k_n) + \nonumber \\
&& \frac{1+r_{n-1} (R_n)^{-1}}{2} \int \{d^2 k_i'\}\, 
W(k_1,k_2,k_3|k_1',k_3') \varphi_{n-1}(k_1',k_3',\cdots,k_n')
\label{compact}
\eeqn
where we use the $(R_n)^{-1}$ operator to show explicitely
that the two last terms in (\ref{collect}) differ only by a cyclic
permutation of the external $n$ gluon indices.

It is therefore possible to require the expression in (\ref{ampuans})
to be an eigenstate of $\bar{K}_n$ if the extra terms, of the form of the last
line in (\ref{compact}), will cancel in the sum, i.e.
\be
\sum_{i=0}^{n-1}  (R_n)^i \Bigl( 1 +r_{n-1} (R_n)^{-1} \Bigr) \, c_i =0,
\ee
which defines the equation for the eigenvalue $r_{n-1}$ with the eigenvectors
components given by the weights $c_i$.
The secular equation is given by $(-r_{n-1})^n=1$, and it must be considered
together with the other constraint $(r_{n-1})^{n-1}=1$, since $r_{n-1}$ has
been previously chosen to be an eigenvalue of the operator $R_{n-1}$.
The solution of this two equations exists and is given by $r_{n-1}=(-1)^n$.

Therefore one obtains the following result: it is possible to construct
$n$-gluon eigenstates using $(n-1)$-gluon eigenstates. It is useful
to distinguish two cases:

(a) $n$ is even; in this case one should use $(n-1)$-gluon eigenstates
    even under rotation ($r_{n-1}=+1$) and weights given by $c_i=(-1)^i$;
    the solution obtained is odd under rotation ($r_n=-1$). Although this is
    a solution of the eq. (\ref{bkpnlarge}), it is not physical. Infact
    by requiring the Bose symmetry the product of the wave functions
    in colour and configuration transverse space should be even under
    rotation over the cylinder while in this case the colour part is even
    (coming from a trace of colour matrices) while the configuration part
    is odd.

    It would be nevertheless interesting to study the possibility
    of the existence of more complicated mixed multi-signature partial waves
    related to a non trivial interplay between transverse
    and longitudinal configuration space, which could allow for a
    physical solution odd under rotation in the tranverse space.

(b) $n$ is odd; in this case the $(n-1)$-gluon eigenstates must be
    odd under rotation ($r_{n-1}=-1$), which means that we start from an
    unphysical solution in the previously mentioned sense
    (for $n > 3$). The weights are given by $c_i=+1$,
    and the solution obtained is even under rotation ($r_n=+1$).
    Therefore it is physical.

It is easy to check that this construction procedure is nilpotent,
in the sense that if from an $(n-1)$-gluon eigenstates an $n$-gluon
eigenstate is constructed, one cannot use the obtained result to construct
an $(n+1)$-gluon state because the result would be identically zero.
The situation is illustrated in figure 1.

\begin{figure}
\label{fig1}
\begin{center}
\input{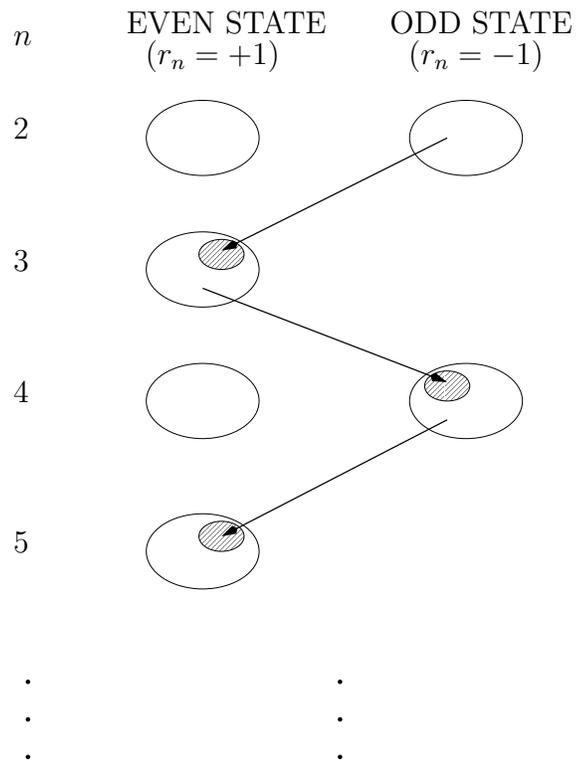}
\caption{An eigenstate of $n$ gluons can be constructed from
an $(n-1)$ gluon eigenstate of opposite parity following the arrows.
The arrows start from a region of functions outside the one pointed
by the incoming arrows, due to the nilpotent property.
Only the states for odd $n$ are physical.}
\end{center}
\end{figure}

If the eigenstates with $(n-1)$ gluons are ``good enough''
the normalization relation for $n$-gluon states can be related to the
normalization of the $(n-1)$-gluon states,
\be
\langle \varphi_n | \varphi_n \rangle =  C_n \frac{2 n}{n-1} \chi \, \,
\langle \varphi_{n-1} | \varphi_{n-1} \rangle,
\label{scalarprod}
\ee
where $C_n$ is a colour factor. In particular 
it is needed that the $(n-1)$-gluon coordinate eigenfunctions are going to
zero fast enough in the limit that two gluon coordinate become the same.

There are case for which this behaviour is not true; for example, for a
function already generated by this construction procedure.
A second iteration is giving
an identically zero eigenfunction due to the nilpotent property and therefore
the norm is zero. On the other hand the amputated function has a $\delta$-like
behaviour in the difference of the coordinates of adiacent gluons and extra
terms will be generated beyond the ones which contribute in eq.
(\ref{scalarprod}), giving totally zero as expected.

We note that such relation will not be valid for eventual
eigenstates with intercept greater than one (the norm would be negative
otherwise), which means that their behaviour, if these eigenstates
exist, has to be more singular than the one indicated above.

The odderon solution \cite{BLV} mentioned at the end of the last section
corresponds to the $n=3$ case.
It is in fact even under cyclic permutation and built by an odd $2$-gluon
(pomeron) eigenstate.

Using the Janik-Wosiek (or other still unknown) odderon solution one could
construct a $4$-gluon state with the same intercept but, as remarked above,
it would be unphysical.
A $5$-gluon state will be therefore, among the states derived in this
framework, the next physical one.

\section{Conclusions} 

It has been shown that a special relation exists between some
solutions of the homogeneous BKP equations for $(n-1)$ and $n$ gluons
in the multicolour limit (\ref{bkpnlarge}). 
In particular it is possible to construct $n$-gluon eigenstates if
$(n-1)$-gluon states are known,
provided they satisfy some symmetry constraints.
These latter ones restrict the set of the physically relevant
solutions which can be constructed to the case of an odd number of gluons.
The recently found family of Odderon states with intercept up to $1$ is
included in this construction.

Of course much more efforts will be necessary to understand the
full complex structure of the solutions of the BKP hierarchy.

It is interesting to note that a few interesting solutions and properties
of such complicated mathematical equations have been suggested by a
very physical relation like the bootstrap for the gluon reggeization.


\section*{Acknowledgements}
The author is very grateful to J. Bartels, L.N. Lipatov and M.A. Braun
for very interesting and helpful discussions.
    
\end{document}